# LiNUS: Lightweight Automatic Segmentation of Deep Brain Nuclei for Real-Time DBS Surgery

Shuo Zhang
*National Engineering Research Center of Neuromodulation*
*Tsinghua University*
Beijing, China
shuo-zha25@mails.tsinghua.edu.cn

Zihua Wang
*National Engineering Research Center of Neuromodulation*
*Tsinghua University*
Beijing, China
wangzh7@mail.tsinghua.edu.cn

Changgeng He
*Software Department*
*Beijing Pins Medical Co., Ltd.*
Beijing, China
hechanggeng815@sina.com

*Chunhua Hu*
*National Engineering Research Center of Neuromodulation*
*Tsinghua University*
Beijing, China
huchunhua@tsinghua.edu.cn
(Corresponding Author)

*Abstract*—This paper proposes LiNUS, a lightweight deep learning framework for the automatic segmentation of the Subthalamic Nucleus (STN) in Deep Brain Stimulation (DBS) surgery. Addressing the challenges of small target volume and class imbalance in MRI data, LiNUS improves upon the U-Net architecture by introducing spectral normalization constraints, bilinear interpolation upsampling, and a multi-scale feature fusion mechanism. Experimental results on the Tsinghua DBS dataset (TT14) demonstrate that LiNUS achieves a Dice coefficient of 0.679 with an inference time of only 0.05 seconds per subject, significantly outperforming traditional manual and registration-based methods. Further validation on high-resolution data confirms the model's robustness, achieving a Dice score of 0.89. A dedicated Graphical User Interface (GUI) was also developed to facilitate real-time clinical application.

## I. INTRODUCTION

Parkinson's disease (PD) is a progressive neurodegenerative disorder characterized by the loss of dopaminergic neurons in the substantia nigra. Deep Brain Stimulation (DBS) has become a standard surgical therapy for advanced PD [1], [2], where high-frequency electrical stimulation is delivered to specific brain targets to modulate abnormal neural circuits. Beyond PD, DBS is also expanding to psychiatric disorders targeting nuclei like the habenula [3]–[6], necessitating precise localization techniques. The Subthalamic Nucleus (STN) is the most common target for DBS (Fig. 1); however, its small volume (approximately $150 - 200\ mm^3$) and low contrast in standard Magnetic Resonance Imaging (MRI) make precise localization a significant challenge.

Accurate electrode placement is critical for surgical success, as misplacement can lead to reduced therapeutic efficacy and adverse side effects. Traditional localization methods rely on the registration of preoperative MRI with brain atlases [7], [8] or manual calculation based on AC-PC coordinates. These methods are labor-intensive, time-consuming (often taking over 10 minutes per case), and prone to inter-observer variability.

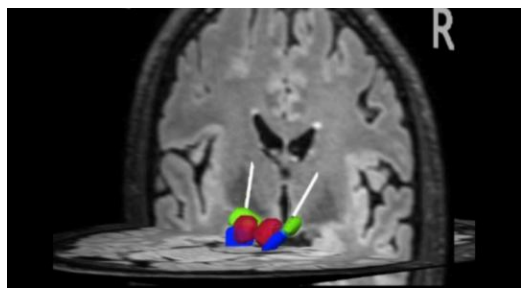

Fig.1. Visualization of brain nucleus localization. The relationship between the target nuclei and electrode trajectories is illustrated in 3D space.

Recent advances in deep learning have shown promise in medical image segmentation. However, existing 3D networks (e.g., V-Net) often suffer from high computational costs and difficulty converging on small, anisotropic datasets. To address these limitations, in this work, we propose LiNUS (Lightweight Nucleus Segmentation), a compact and efficient 2D network designed for the automatic segmentation of deep brain nuclei. By integrating spectral normalization and replacing transposed convolutions with bilinear interpolation, LiNUS achieves a balance between high segmentation accuracy and real-time processing speed suitable for intraoperative navigation.

## II. METHODOLOGY

*A. Datasets*

Two datasets were utilized in this study:

- **TT14 (Tsinghua DBS Dataset):** This dataset includes 3D T1-weighted MRI scans from 14 patients with primary Parkinson's disease (Table I). Representative data samples are shown in Fig. 2. The images have a resolution of $1.0 \times 1.0 \times 1.0\ mm^3$ and dimensions of $182 \times 218 \times 182$ voxels.

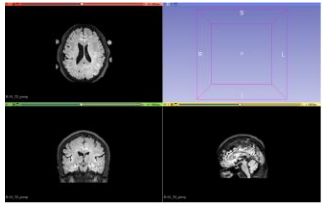

Fig.2. Three-view schematic of the MRI dataset utilized in the experiments, displaying axial, coronal, and sagittal perspectives.

- **PD25 Dataset:** A high-resolution multi-contrast average brain atlas constructed from 25 PD patients [9], [10], with a voxel size of $0.3 \times 0.3 \times 0.3\ mm^3$. This dataset was used to validate the impact of image quality on segmentation performance.

TABLE I. THE DEMOGRAPHIC AND CLINICAL INFORMATION OF PATIENTS IN THE TT14 DATASET

| ID | Age | Sex | Edu. (y) | MMSE | H-Y | Dis.Dur. (y) | UPDRS-III | PDQ39 |
|---|---|---|---|---|---|---|---|---|
| DBS01 | 46 | M | 9 | 26 | 3 | 8 | 22 | 75 |
| DBS02 | 53 | M | 12 | 25 | 3 | 13 | 34 | 34 |
| DBS03 | 67 | M | 12 | 27 | 3 | 8 | 21 | 78 |
| DBS04 | 51 | M | 12 | 27 | 4 | 20 | 20 | 40 |
| DBS05 | 65 | M | 12 | 28 | 3 | 6 | 21 | 14 |
| DBS06 | 60 | F | 9 | 28 | 3 | 7 | 29 | 71 |
| DBS07 | 46 | M | 15 | 29 | 4 | 7 | 47 | 60 |
| DBS08 | 59 | M | 6 | 28 | 4 | 8 | 29 | 51 |
| DBS09 | 61 | M | 6 | 28 | 4 | 8 | 29 | 51 |
| DBS10 | 51 | M | 15 | 28 | 1.5 | 6 | 8 | 75 |
| DBS11 | 47 | F | 9 | 28 | 5 | 12 | 38 | 68 |
| DBS12 | 56 | F | 15 | 26 | 4 | 15 | 51 | 61 |
| DBS13 | 61 | F | 9 | 30 | 4 | 8 | 41 | 127 |
| DBS14 | 51 | M | 12 | 30 | 4 | 8 | 54 | 39 |
| Mean | 54.9 | | 10.9 | 27.7 | 3.5 | 9.7 | 32.3 | 60.3 |
| Variance | 7.7 | | 3.0 | 1.4 | 0.8 | 4.0 | 13.3 | 26.7 |

## B. LiNUS Architecture

LiNUS is built upon an Encoder-Decoder structure with skip connections, specifically modified to handle the small-target nature of the STN. The network architecture is illustrated in Fig. 3.

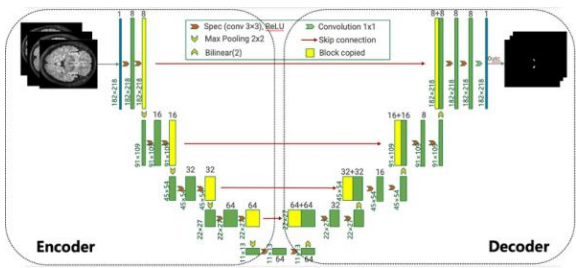

Fig.3. Schematic diagram of the LiNUS network architecture.

*1) Encoder with Spectral Normalization:* The encoder processes 2D MRI slices to extract high-level semantic features. It consists of four downsampling stages. Each stage employs a "DoubleConv" module (two $3 \times 3$ convolutions, ReLU activation) followed by Spectral Normalization [11] (an improvement over batch normalization [12]). Spectral normalization constrains the Lipschitz constant of the network weights, preventing gradient explosion and enhancing training stability, which is crucial for small datasets.

*2) Decoder with Bilinear Interpolation:* Traditional U-Net architectures use transposed convolutions for upsampling, which can introduce checkerboard artifacts. LiNUS replaces these with bilinear interpolation, which generates smoother object boundaries. Our experiments indicated that this modification improved the Dice coefficient at the STN boundary by approximately 0.02.

*3) Skip Connections:* To recover spatial information lost during downsampling, skip connections concatenate feature maps from the encoder with the corresponding upsampled features in the decoder, fusing shallow textural details with deep semantic information.

## C. Training Strategy

The model was trained using a weighted hybrid loss function combining Dice Loss and Cross-Entropy Loss to address the extreme class imbalance between the small STN and the background .

- **Hyperparameters:** Batch size = 128; Epochs = 200.
- **Optimization:** A Cosine Annealing learning rate schedule was applied, decaying from 0.001 to 0.0001, to ensure stable convergence.
- **Augmentation:** Extensive data augmentation (rotation, scaling, elastic deformation, Gaussian noise) was performed to prevent overfitting.

## III. EXPERIMENTAL RESULTSY

### A. Segmentation Performance

We compared LiNUS against six state-of-the-art neural networks (Lraspp [13], V-Net [14], U-Net [15], U-Net3D [16], FCN [17], Deeplab[18]) on the TT14 dataset. As detailed in Table II, LiNUS demonstrated superior performance in terms of accuracy and efficiency.

TABLE II. PERFORMANCE COMPARISON OF LINUS WITH OTHER NETWORKS

| Model | Dice | Precision | Recall | Size (bytes) | Params. | Time (s) |
|---|---|---|---|---|---|---|
| LiNUS | 0.6790 | 0.6808 | 0.7590 | 890792 | 222698 | 0.05 |
| Lraspp | 0.4241 | 0.3920 | 0.6000 | 13082225 | 3270557 | 4.04 |
| VNet | 0.0001 | 0.0001 | 1.0000 | 90286028 | 22571507 | 36.4 |
| UNet | 0.5501 | 0.7047 | 0.4901 | 97896390 | 24474098 | 4.55 |
| UNet3d | 0.0001 | 0.0000 | 0.0998 | 1223078 | 305770 | 24.27 |

| Model | Dice | Precision | Recall | Size (bytes) | Params. | Time (s) |
|---|---|---|---|---|---|---|
| FCN | 0.5103 | 0.4120 | 0.7785 | 132099198 | 33024800 | 6.07 |
| Deeplab | 0.4664 | 0.3514 | 0.8463 | 158888798 | 39722200 | 12.13 |

LiNUS achieved the highest Dice coefficient of 0.6790, significantly outperforming the standard U-Net (0.5501) and FCN (0.5103). Notably, 3D networks like V-Net and U-Net3D failed to converge effectively on this dataset (Dice 0.0001), likely due to the limited sample size and anisotropy of the clinical data. In terms of efficiency, LiNUS is extremely lightweight, with only 0.89 million parameters. The average inference time for a single patient (processing all slices) is 0.05 seconds (Table III), which is orders of magnitude faster than traditional registration methods (>10 minutes) and significantly faster than other deep learning models.

TABLE III. EFFICIENCY COMPARISON: LINUS VS. TRADITIONAL METHODS

| Method | Dice | Execution time(s) |
|---|---|---|
| LiNUS | 0.6790 | 0.05s |
| Traditional method | 0.68 | 15min |

### B. Localization Accuracy

To evaluate the clinical utility, we calculated the Euclidean distance between the predicted STN center of mass and the ground truth. The mean localization error was approximately 1.699 mm, as detailed in Table IV and Table V. Given that the STN length ranges from 9 to 14 mm, this error margin is considered acceptable for preoperative planning support.

TABLE IV. LOCALIZATION ERROR ANALYSIS FOR LEFT STN (PHYSICAL DISTANCE)

| Prediction | Ground Truth | Physical Distance(mm) |
|---|---|---|
| {61,76,104} | {61,76,104} | 0 |
| {64,79,107} | {63,79,111} | 4.12 |
| {65,75,108} | {65,75,108} | 0 |
| {62,73,107} | {62,73,110} | 3 |
| {67,74,111} | {67,80,113} | 6.32 |
| {64,75,110} | {60,78,109} | 5.10 |
| {60,97,113} | {61,102,108} | 7.14 |
| {67,74,108} | {67,74,108} | 0 |
| {62,76,108} | {63,76,108} | 1 |
| {67,75,105} | {67,75,105} | 0 |
| {67,74,109} | {67,74,109} | 0 |
| {63,104,108} | {63,105,108} | 1 |
| {66,74,105} | {67,74,106} | 1.41 |
| {63,74,112} | {63,74,112} | 0 |

TABLE V. LOCALIZATION ERROR ANALYSIS FOR RIGHT STN (PHYSICAL DISTANCE)

| Prediction | Ground Truth | Physical Distance(mm) |
|---|---|---|
| {60,99,111} | {60,99,110} | 1 |
| {62,96,111} | {63,96,109} | 2.24 |
| {62,98,111} | {63,98,112} | 1.41 |
| {60,98,109} | {60,98,109} | 0 |
| {65,99,112} | {65,99,112} | 0 |
| {62,99,113} | {62,99,113} | 0 |
| {60,98,112} | {62,90,119} | 10.82 |
| {62,100,110} | {62,100,110} | 0 |
| {64,101,112} | {63,101,112} | 1 |
| {63,98,109} | {63,98,109} | 0 |
| {61,99,109} | {61,99,110} | 1 |
| {62,100,110} | {62,101,110} | 1 |
| {63,98,109} | {63,98,109} | 0 |
| {62,99,114} | {62,99,114} | 0 |

### C. Generalization and Data Quality

The model's performance is heavily influenced by image resolution. When trained and tested on the high-resolution PD25 dataset (Table VI), LiNUS's Dice coefficient improved to 0.895812 (Table VII). This suggests that the architecture is capable of high-precision segmentation given sufficient image quality. Furthermore, by incorporating the MI-ST optimizer [19], the framework was successfully extended to segment the Globus Pallidus internus (GPi) with a Dice score of 0.9139 (Table VIII), demonstrating strong generalization capabilities. Additionally, comparative experiments (Table IX) indicate that the MI-ST optimizer outperforms the traditional Adam optimizer [20], ensuring better stability and higher segmentation accuracy in these complex subcortical tasks.

TABLE VI. COMPARISON OF IMAGE PARAMETERS BETWEEN TT14 AND PD25 DATASETS

| Dataset | Image dimensions | Image spacing |
|---|---|---|
| TT14 | 182*182*182 | 1*1*1mm |
| PD25 | 354*334*334 | 0.3*0.3*0.3mm |

TABLE VII. PERFORMANCE IMPROVEMENT OF LINUS ON THE HIGH-RESOLUTION PD25 DATASET

| Dataset | Dice | Precision | Recall | Execution time(s) |
|---|---|---|---|---|
| TT14 | 0.6790 | 0.6808 | 0.7590 | 0.05 |
| PD25 | 0.8958 | 0.8370 | 0.9930 | 0.05 |

TABLE VIII. GENERALIZATION TEST: GPI VS. STN LOCALIZATION ON PD25 DATASET

| Dataset | Dice | Precision | Recall | Execution time(s) |
|---|---|---|---|---|
| PD25(GPi) | 0.9139 | 0.9010 | 0.8790 | 0.05 |
| PD25(STN) | 0.8958 | 0.8370 | 0.9930 | 0.05 |

TABLE IX. IMPACT OF DIFFERENT OPTIMIZERS (MI-ST VS. ADAM) ON SEGMENTATION DICE SCORES

| Optimizer | Dice |
|---|---|
| MI-ST | 0.9179 |
| Adam | 0.9139 |

## IV. DISCUSSION

### A. Speed vs. Accuracy

A key contribution of this work is the balance between speed and accuracy. While traditional methods and heavy 3D networks incur high computational costs, LiNUS's lightweight design allows for near-instantaneous feedback (0.05s). This speed is critical for intraoperative scenarios where plans may need to be updated in real-time.

### B. The "Grokking" Phenomenon

During training, we observed a "grokking" phenomenon (Fig. 4) where the model's accuracy remained near zero for an extended period before suddenly converging around epoch 170 [21]. This highlights the importance of prolonged training schedules and appropriate regularization when training on small, imbalanced medical datasets.

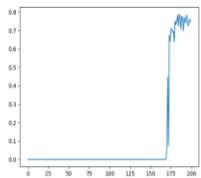

Fig.4. The "grokking" phenomenon observed during training: accuracy remains low for distinct epochs before sudden convergence.

### C. Limitations

A current limitation is the use of 2D slices, which ignores spatial continuity along the Z-axis, potentially leading to minor discontinuities in the 3D reconstruction (Fig. 5, Fig. 6). Future work will explore 2.5D input strategies (stacking adjacent slices) or post-processing with Total Variation loss to enforce 3D smoothness without the computational overhead of fully 3D networks

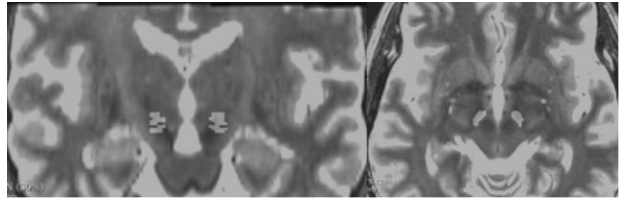

Fig.5. Illustration of spatial discontinuity issues observed in 3D MRI reconstruction from 2D slice predictions.

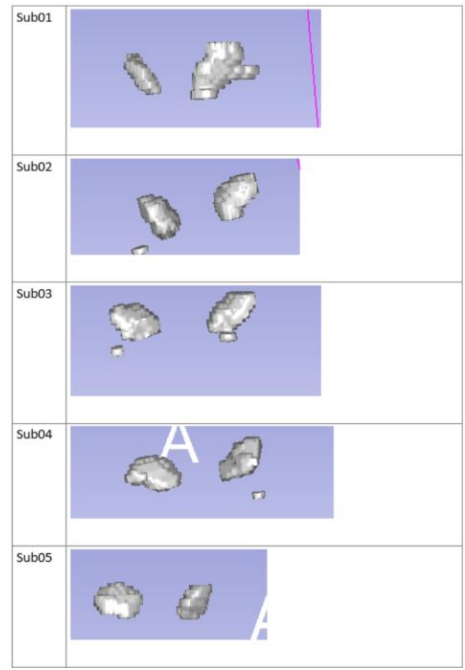

Fig.6. Visualization of discrete segmentation artifacts across different subjects due to lack of Z-axis context.

## V. GRAPHICAL USER INTERFACE

To facilitate clinical application, we developed a dedicated software system based on the proposed LiNUS. As illustrated in Fig. 7, the system workflow integrates data preprocessing, model inference, and 3D visualization into a streamlined pipeline. The user interface (Fig. 8) allows surgeons to load patient MRI data and obtain real-time segmentation masks instantly.

The system provides multi-view visualization capabilities (Fig. 9), enabling simultaneous inspection of the target nuclei from axial, coronal, and sagittal perspectives.

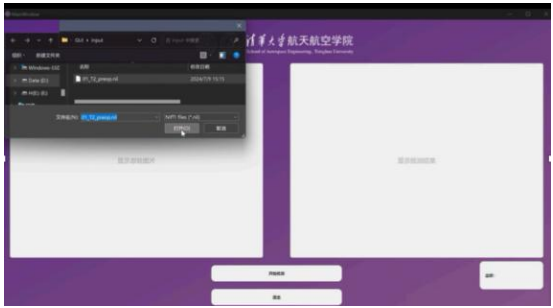

Fig.7.  Schematic diagram of the LiNUS GUI.

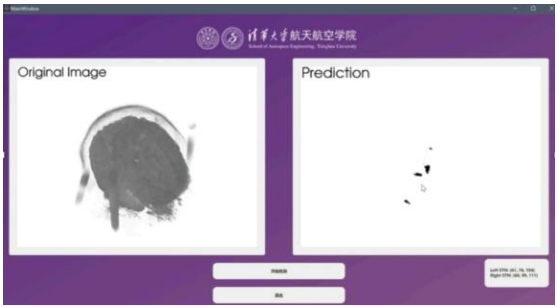

Fig.8.  The GUI interface displaying original MRI input and real-time prediction masks.

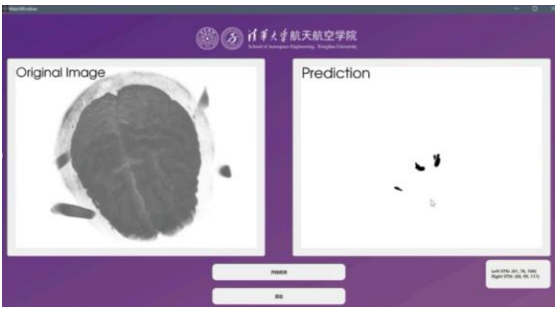

Fig.9.  Multi-view visualization (Axial, Coronal, Sagittal) in the developed GUI.

## VI. CONCLUSION

This paper presents LiNUS, a rapid and accurate deep learning framework tailored for STN localization in DBS surgery. By addressing the challenges of small target volume and data anisotropy through spectral normalization and bilinear interpolation, LiNUS achieves state-of-the-art segmentation performance while maintaining extremely low computational cost. The proposed method significantly outperforms traditional registration-based approaches in terms of inference speed (0.05s per subject), making it suitable for intraoperative scenarios.

Validation on both the TT14 and high-resolution PD25 datasets confirms the model's robustness and generalization capability. Furthermore, the successful integration of the model into a user-friendly GUI bridges the gap between algorithm development and clinical practice. Future work will focus on incorporating multi-modal data fusion (e.g., fusing MRI with CT) and developing "human-in-the-loop" interactive mechanisms to further enhance safety and reliability in surgical decision-making.